\documentclass[a4paper,12pt]{article}
\usepackage{a4wide}
\usepackage[latin1]{inputenc}
\usepackage{dsfont}
\usepackage{amsfonts}
\usepackage{amsmath}
\usepackage{amssymb}
\usepackage{amsthm}
\usepackage{graphicx}
\usepackage{booktabs}
\usepackage[numbers, sort]{natbib}
\allowdisplaybreaks[1]
\usepackage{color}
\usepackage{ulem}
\usepackage{tikz}
\usetikzlibrary{arrows}

\usepackage[font=footnotesize,bf]{caption}
\newcommand{\be}{\begin{equation}}
\newcommand{\ee}{\end{equation}}
\newcommand{\beq}{\begin{equation}}
\newcommand{\eeq}{\end{equation}}
\newcommand{\bea}{\begin{eqnarray}}
\newcommand{\eea}{\end{eqnarray}}

\DeclareMathOperator{\diag}{diag}

\def\nuL{\nu}
\def\EL{\mathrm{e}}

\begin{document}

\begin{titlepage}

\vspace*{-15mm}

\vspace*{0.7cm}

\begin{center}
{\Large {\bf $\theta^{\text{\text{PMNS}}}_{13} = \theta_C / \sqrt{2}$ from GUTs 
}}\\[8mm]

Stefan Antusch$^{\star\dagger}$\footnote{Email: \texttt{stefan.antusch@unibas.ch}},~
Christian Gross$^{\star}$\footnote{Email: \texttt{christian.gross@unibas.ch}},~
Vinzenz Maurer$^{\star}$\footnote{Email: \texttt{vinzenz.maurer@unibas.ch}},~
Constantin Sluka$^{\star}$\footnote{Email: \texttt{constantin.sluka@unibas.ch}}
\end{center}
\addtocounter{footnote}{-4}

\vspace*{0.20cm}

\centerline{$^{\star}$ \it
Department of Physics, University of Basel,}
\centerline{\it
Klingelbergstr.\ 82, CH-4056 Basel, Switzerland}

\vspace*{0.4cm}

\centerline{$^{\dagger}$ \it
Max-Planck-Institut f\"ur Physik (Werner-Heisenberg-Institut),}
\centerline{\it
F\"ohringer Ring 6, D-80805 M\"unchen, Germany}

\vspace*{1.2cm}

\begin{abstract}
\noindent
The recent observations of the leptonic mixing angle $\theta^{\text{PMNS}}_{13}$ are consistent with $\theta^{\text{PMNS}}_{13} = \theta_C / \sqrt{2}$ (with $\theta_C$ being the Cabibbo angle $\theta^{\text{CKM}}_{12}$). We discuss how this relation can emerge in Grand Unified Theories (GUTs) via charged lepton corrections. The key ingredient is that in GUTs the down-type quark Yukawa matrix and the charged lepton Yukawa matrix are generated from the same set of GUT operators, which implies that the resulting entries are linked and differ only by group-theoretical Clebsch factors. This allows a link $\theta^e_{12} \approx \theta_C$ to be established, which can induce $\theta^{\text{PMNS}}_{13} \approx \theta_C / \sqrt{2}$ provided that the 1-3 mixing in the neutrino mass matrix is much smaller than $\theta_C$. We find simple conditions under which $\theta^{\text{PMNS}}_{13} \approx \theta_C / \sqrt{2}$ can arise via this link in SU(5) GUTs and Pati-Salam models. We also discuss possible corrections to the exact relation.
Using lepton mixing sum rules different neutrino mixing patterns can be distinguished by their predictions for the Dirac CP phase $\delta^{\text{PMNS}}$.  
\end{abstract}

\end{titlepage}

\newpage

\section{Introduction}
The question of the origin of the quark and lepton masses and mixing parameters is one of the biggest unsolved problems in particle physics. In addition to the smallness of the neutrino masses, the largeness of their mixing compared to the mixing angles in the quark sector provides a major puzzle.

While the quark mixing angles, in the standard PDG  parametrisation for the \text{\text{CKM}} matrix $U_{\text{CKM}}$, are about (at 1$\sigma$) \cite{pdg}
\be
\theta^{\text{CKM}}_{13} = 0.20^\circ \pm 0.01^\circ  \; , \quad \theta^{\text{CKM}}_{23} = 2.35^\circ \pm 0.07^\circ\; , \quad \theta^{\text{CKM}}_{12} \equiv \theta_C =  13.02^\circ \pm 0.04^\circ \;,
\ee
it has been known for some time that two of the leptonic mixing angles, $\theta^{\text{PMNS}}_{23}$ and $\theta^{\text{PMNS}}_{12}$ in the standard PDG parametrisation for the \text{\text{PMNS}} matrix $U_{\text{PMNS}}$, are large. 
Recently, the third leptonic mixing angle $\theta^{\text{PMNS}}_{13}$ has been measured by T2K \cite{t2k}, DoubleCHOOZ \cite{Abe:2011fz}, DayaBay \cite{An:2012eh} and RENO \cite{Ahn:2012nd}, revealing, when combined with the results from global fits \cite{Schwetz:2011zk}\footnote{We note that the global fit of \cite{Fogli:2011qn} finds a best fit value of $\theta_{23}^\text{PMNS}=40.4^\circ$ with a one sigma range of $[38.6^\circ,45.0^\circ]$. We come back to this in section \ref{sec:corr}.}, the following mixing pattern in the lepton sector (at 1$\sigma$):
\be
\theta^{\text{PMNS}}_{13} = 8.8^\circ \pm 1.0^\circ  \; , \quad \theta^{\text{PMNS}}_{23} = 46.1^\circ\pm 3.4^\circ \; , \quad \theta^{\text{PMNS}}_{12} = 34.0^\circ \pm  1.1^\circ  \;.
\ee
The experimental result of this not-so-small $\theta^{\text{PMNS}}_{13}$ has triggered a large interest in the community (see e.g.~\cite{bulk}).

Given the rather precise measurement of $\theta^{\text{PMNS}}_{13}$, one may ask the question whether the fact that $\theta^{\text{PMNS}}_{13}$ numerically agrees well with
\be
\label{Eq:3}
\theta^{\text{PMNS}}_{13} = \frac{\theta_C}{\sqrt{2}} \approx 9.2^\circ
\ee
 has any particular meaning. In particular, one may wonder whether this specific connection to $\theta_C$ can be a consequence of an underlying Grand Unified Theory (GUT). In GUTs, quarks and leptons are unified in joint representations of the GUT symmetry group, which implies that the flavour structures in the quark and the lepton sectors are linked and may lead to the appearance of $\theta_C$ in Eq.~(\ref{Eq:3}).
 
While this possibility is certainly attractive, it should be noted that many existing GUT models have predicted smaller values of $\theta^{\text{PMNS}}_{13}$ (often around $3^\circ$), especially when they were using the Georgi-Jarlskog Clebsch factor of $3$ to obtain viable mass relations for the first two quark and lepton families \cite{GJ}. Only recently, alternative (combinations of) Clebsch factors have been proposed \cite{Antusch:2009gu}, which can lead to viable mass relations and comparatively large values of $\theta^{\text{PMNS}}_{13}$ \cite{Antusch:2011qg,Marzocca:2011dh}, in accordance with the experimental hints at this time. In the light of the recent rather precise measurement, which suggests Eq.~(\ref{Eq:3}) as a possible GUT connection, and given the new possibilities for model building, it is interesting to analyze how the relation of Eq.~(\ref{Eq:3}) may actually be realized in GUTs.    

We would like to note that even before the recent $\theta_{13}^\text{PMNS}$-measurements various different approaches featuring large values of $\theta_{13}^\text{PMNS}$ existed in the literature, as we will now briefly discuss. For example, $\theta_{13}^\text{PMNS} = {\cal O} (\theta_C)$ from charged lepton corrections is generally expected if the Wolfenstein parameter $\lambda \approx \theta_C$ is assumed as an expansion parameter in the mass matrices of the quark sector as well as of the lepton sector, as in scenarios with ``Cabibbo haze'' \cite{Datta:2005ci}, or e.g.\ in \cite{cabbibobulk}. However, $\theta_{13}^\text{PMNS}$ differs from $\lambda$ by an unknown prefactor, which means its value is not predicted. 
More specific relations, among them also $\theta_{13}^\text{PMNS} = \theta_C/\sqrt{2}$, have appeared in the context of Quark-Lepton Complementarity (QLC) \cite{Minakata:2004xt,QLC,Patel:2010hr}, which was motivated by the possibility that $\theta_{12}^\text{PMNS} + \theta_C = 45^\circ$. 

In the above-mentioned version of QLC \cite{Minakata:2004xt}, the relation $\theta_{13}^\text{PMNS} \approx \theta_C/\sqrt{2}$ arises when one assumes that the neutrino mixing is ``bi-maximal'' \cite{Barger:1998ta} and that the charged lepton mixing matrix is exactly equal to the CKM mixing matrix. The latter assumption, however, is not expected to hold in realistic GUTs since the mass matrices of quarks and charged leptons cannot be identical, especially regarding the masses of the first two families. In particular, the mass ratio $m_\mu / m_s$ at the GUT scale clearly differs from $1$, and therefore non-universal group-theoretical Clebsch factors are required, as mentioned above. Since the mass eigenvalues differ, it is not expected that the charged lepton mixing matrix equals the CKM matrix. In our approach, we will therefore neither require ``bi-maximal'' mixing in the neutrino sector nor full equality between the charged lepton mixing matrix and the CKM matrix. We will come back to a comparison to QLC in section \ref{sec:sumrule}.

 The goal of this paper is to identify simple conditions under which GUT models can give rise to $\theta^{\text{PMNS}}_{13} \approx \theta_C / \sqrt{2}$. The strategy is to use GUT relations plus additional conditions/constraints on the structure of SU(5) GUTs and Pati-Salam models to establish $\theta^e_{12} \approx \theta_C$, which then induces $\theta^{\text{PMNS}}_{13} \approx \theta_C / \sqrt{2}$ via this charged lepton mixing contribution. We also emphasize that, although we are aiming at explaining $\theta_{13}^\text{PMNS} \approx \theta_C/ \sqrt{2}$ beyond $\theta_{13}^\text{PMNS} = \mathcal{O}(\theta_C)$, the relation can never be exact in a realistic model. We therefore discuss various possible corrections to $\theta_{13}^\text{PMNS} =\theta_C/ \sqrt{2}$ (cf.\ section \ref{sec:corr}) and find that they are typically expected to be less than about 10\%, i.e., of the order of the present experimental uncertainty. 
For a given specific model, the deviation from the exact relation can be predicted, and future more precise measurements of $\theta_{13}^\text{PMNS}$ (and $\theta_{23}^\text{PMNS}$) may allow to discriminate between different models realizing  $\theta_{13}^\text{PMNS} \approx \theta_C / \sqrt{2}$.

\section{Conditions from relations between $Y_u$ and $Y_d$ and between $m_\nu$ and $Y_e$}
\label{sec:1}
In the following, we will assume hierarchical $Y_u, Y_d$ and $Y_e$, which implies that the left-mixing angles (named $\theta^u_{ij}, \theta^d_{ij}$ and $\theta^e_{ij}$) are all comparatively small (i.e., not much larger than the Cabibbo angle $\theta_C$) as is typical for GUT flavour models in the flavour basis. To find simple conditions\footnote{We do not claim that the conditions discussed in the following are the only ones which can lead to $\theta^{\text{PMNS}}_{13} \approx \theta_C / \sqrt{2}$.} for obtaining a predictive GUT scenario with $\theta^{\text{PMNS}}_{13} \approx \theta_C / \sqrt{2}$, we first revisit the relations between the mixing parameters in the quark and lepton sectors. We choose the standard-parametrization for a general unitary matrix $U$:
\begin{eqnarray}\label{eq:StandardParametrizationU}
 U & = & \diag(e^{i\delta_e},e^{i\delta_\mu},e^{i\delta_\tau})\cdot V \cdot 
 \diag(e^{-i\varphi_1/2},e^{-i\varphi_2/2},1)
\end{eqnarray}
where 
\begin{equation}
 V=\left(
 \begin{array}{ccc}
 c_{12}c_{13} & s_{12}c_{13} & s_{13}e^{-i\delta}\\
 -c_{23}s_{12}-s_{23}s_{13}c_{12}e^{i\delta} &
 c_{23}c_{12}-s_{23}s_{13}s_{12}e^{i\delta} & s_{23}c_{13}\\
 s_{23}s_{12}-c_{23}s_{13}c_{12}e^{i\delta} &
 -s_{23}c_{12}-c_{23}s_{13}s_{12}e^{i\delta} & c_{23}c_{13}
 \end{array}
 \right)
\end{equation}
with \(c_{ij}\) and \(s_{ij}\) defined as \(\cos\theta_{ij}\) and
\(\sin\theta_{ij}\), respectively. 
Starting with the quark sector and expanding to leading order in the small mixing angles,
we obtain the following relations~\cite{Antusch:2009hq}:
\begin{subequations}
\bea \label{F1}
{\theta^{\text{CKM}}_{23}}e^{-i\delta^{\text{CKM}}_{23}}&=&
{\theta_{23}^{d}}e^{-i\delta_{23}^{d}}
-{\theta_{23}^{u}}e^{-i\delta_{23}^{u}}\;,
\\
\label{F2} {\theta^{\text{CKM}}_{13}}e^{-i\delta^{\text{CKM}}_{13}}&=&
\theta_{13}^{d}e^{-i\delta_{13}^{d}}-\theta_{13}^{u}e^{-i\delta_{13}^{u}}-{\theta_{12}^{u}}e^{-i\delta_{12}^{u}}
({\theta_{23}^{d}}e^{-i\delta_{23}^{d}} - {\theta_{23}^{u}}e^{-i\delta_{23}^{u}})
\;,\\
\label{F3} {\theta^{\text{CKM}}_{12}}e^{-i\delta^{\text{CKM}}_{12}}&=&
{\theta_{12}^{d}}e^{-i\delta_{12}^{d}}
-{\theta_{12}^{u}}e^{-i\delta_{12}^{u}} \;.
\eea
\end{subequations}
The standard PDG Dirac CP phase $\delta_{\rm \text{CKM}}$ in $U_{\text{CKM}}$ can be identified as (cf.~\cite{King:2002nf})
\be\label{eq:deltafromparam1}
\delta_{\text{CKM}} = \delta^{\text{CKM}}_{13}-\delta^{\text{CKM}}_{23}-\delta^{\text{CKM}}_{12} \;,
\ee
while the phases $\delta^{u}_{ij}$ and $\delta^{d}_{ij}$ are associated with the rotation angles $\theta^{u}_{ij}$ and $\theta^{d}_{ij}$ in the up- and down-quark sectors, using the same notation as in \cite{Antusch:2009hq}.

Similar considerations can be done in the lepton sector, where we obtain in leading order in a small angle expansion  (treating also $\theta^{\text{PMNS}}_{13}$ as a ``small'' parameter) \cite{Antusch:2005kw}:
 \begin{subequations}\bea 
\label{Eq:23} s^{\text{PMNS}}_{23}e^{-i\delta^{\text{PMNS}}_{23}}
& = &
s_{23}^{\nuL}e^{-i\delta_{23}^{\nuL}}
-\theta_{23}^{\EL}
c_{23}^{\nuL}e^{-i\delta_{23}^{\EL}}\;,
\label{chlep23}
\\
\label{Eq:13} \theta^{\text{PMNS}}_{13}e^{-i\delta^{\text{PMNS}}_{13}}
& = &
\theta_{13}^{\nuL}e^{-i\delta_{13}^{\nuL}}
-\theta_{13}^{\EL}c_{23}^{\nuL}e^{-i\delta_{13}^{\EL}}
-\theta_{12}^{\EL}e^{-i\delta_{12}^{\EL}}(s_{23}^{\nuL}e^{-i\delta_{23}^{\nuL}}-\theta_{23}^{\EL}e^{-i\delta_{23}^{\EL}})\;,
\label{chlep13}
\\
\label{Eq:12} s^{\text{PMNS}}_{12}e^{-i\delta^{\text{PMNS}}_{12}}
& = &
s_{12}^{\nuL}e^{-i\delta_{12}^{\nuL}}
+\theta_{13}^{\EL}
c_{12}^{\nuL}s_{23}^{\nuL}e^{i(\delta_{23}^{\nuL}-\delta_{13}^{\EL})}
- \theta_{12}^{\EL} c_{23}^{\nuL}c_{12}^{\nuL}e^{-i\delta_{12}^{\EL}}\;.
\label{chlep12}
\eea\end{subequations}
The leptonic CP phases can be extracted via
\be\label{Eq:Dict}
\delta^{\text{PMNS}}_{23} = -\frac{\varphi^{\text{PMNS}}_2}{2} \;,\quad
\delta^{\text{PMNS}}_{13} = \delta^{\text{PMNS}} - \frac{\varphi^{\text{PMNS}}_1}{2} \;,\quad
\delta^{\text{PMNS}}_{12} = \frac{1}{2}\left(\varphi^{\text{PMNS}}_2 - \varphi^{\text{PMNS}}_1\right) \; ,
\ee
where $\varphi^{\text{PMNS}}_1$ and $\varphi^{\text{PMNS}}_2$ are the Majorana phases and $\delta^{\text{PMNS}}$ is the Dirac CP phase. We have also included the term of order $\theta_{12}^e\theta^e_{23}$ in Eq.~(\ref{chlep13}) to show that the measured angle $\theta_{23}^\text{PMNS}$ appears in the prefactor of the $\theta_{12}^{\EL}$ term of Eq.~(8b):
\be
\theta^{\text{PMNS}}_{13}e^{-i\delta^{\text{PMNS}}_{13}}
 = 
\theta_{13}^{\nuL}e^{-i\delta_{13}^{\nuL}}
-\theta_{13}^{\EL}c_{23}^{\nuL}e^{-i\delta_{13}^{\EL}}
-\theta_{12}^{\EL}s^{\text{PMNS}}_{23}e^{-i(\delta^{\text{PMNS}}_{23}+\delta_{12}^{\EL})}\;.
\label{chlep13PMNS23}
\ee

We are now ready to state first conditions towards scenarios featuring $\theta^{\text{PMNS}}_{13} \approx \theta_C / \sqrt{2}$: 
\begin{description}

\item[Condition 1:] Since we want to use the charged lepton correction proportional to $\theta_{12}^{\EL}$ to establish the link to GUTs we require, as assumed to derive Eqs.~(6), (8) and (10), hierarchical $Y_u$, $Y_d$ and $Y_e$, and furthermore 
\be
\theta_{13}^\nu \approx 0 \; , \quad \theta_{13}^e \approx 0\; .
\ee
Then the first two summands on the right side of Eq.~(\ref{chlep13PMNS23}) drop out and we obtain, independently of any phases, 
\be\label{Eq:t13fromt12e}
\theta^{\text{PMNS}}_{13} \approx \theta_{12}^{\EL}s_{23}^\text{PMNS} \approx \frac{1}{\sqrt{2}} \theta_{12}^{\EL} \;.
\ee
In the last step, we have inserted a maximal atmospheric mixing angle (i.e.\ $\theta^{\text{PMNS}}_{23} = 45^\circ$) to keep the discussion simple. Since the current experimental value of $s_{23}^{\text{PMNS}}$ is $0.72\pm0.04$ \cite{Schwetz:2011zk}, an uncertainty of approximately $6\%$ is introduced in Eq.~(\ref{Eq:t13fromt12e}). 

\item[Condition 2:] Secondly, since we want to establish a link between $\theta^{\text{PMNS}}_{13}$ and $\theta_C \equiv \theta^{\text{CKM}}_{12}$ based on GUT relations between $Y_d$ and $Y_e$, we require that 
\be
\theta^d_{12} = \theta_C 
\ee
to a good approximation. This may be a consequence of $\theta_{12}^u\ll\theta_{12}^d$, which is a typical feature of models with hierarchical Yukawa matrices where the stronger hierarchy in the up-quark sector implies the smaller mixing angles. An alternative possibility arises when one requires $\theta_{13}^u$, $\theta_{13}^d\approx 0$, which leads to $\theta_{12}^d\approx 12.0^\circ\pm0.3^\circ$ via the quark mixing sum rule $\theta_{12}^d \approx \Big |\theta_{12}^\text{CKM}-\frac{\theta_{13}^\text{CKM}}{\theta_{23}^\text{CKM}}e^{-i\delta^\text{CKM}}\Big |$ (cf.~\cite{Antusch:2009hq}).

\end{description}

\section{Conditions from relations between $Y_e$ and $Y_d$}
\label{sec:2}
To arrive at $\theta^{\text{PMNS}}_{13} \approx \theta_C / \sqrt{2}$, we want to make use of GUT relations between the down-type Yukawa matrix $Y_d$ and the charged lepton Yukawa matrix $Y_e$. The GUT relations emerge since the down-type quark Yukawa matrix and the charged lepton Yukawa matrix are generated from joint GUT operators. 
This leads to the following condition:

\begin{description}
\item[Condition 3:] To obtain predictive GUT relations, we require that the elements of the Yukawa matrices $Y_e$ and $Y_d$ are each dominantly generated by one single joint GUT operator.\footnote{We note that this condition can be somewhat relaxed in specific cases. For instance, not all matrix elements but only the ones which enter the relation between $\theta_C$ and $\theta_{12}^e$ are subject to this requirement. Furthermore, it may also be that two operators featuring the same Clebsch factor contribute at similar strength. In this case, the relation between the elements of $Y_d$ and $Y_e$ is still given by only one Clebsch factor. We make this somewhat stronger assumption at this point to keep the discussion simple.} When this is the case, the matrix elements are closely linked by group theoretical Clebsch factors. 
Focusing now explicitly on the 1-2 submatrix of the first two families, which is a good approximation since we are assuming small mixings in $Y_e$ and $Y_d$, we can write
\be\label{Eq:YevsYd}
Y_d = \begin{pmatrix} 
d&b\\
a&c
\end{pmatrix}\:
\Rightarrow 
\quad
\begin{cases}
\mbox{In PS:} \;\: &  Y_e = 
\begin{pmatrix} 
c_d \,d& c_b \, b\\
 c_a\, a&c_c \, c
\end{pmatrix}\;, \\[6mm]
\mbox{In SU(5):}\;\:&  Y_e = 
\begin{pmatrix} 
c_d \,d& c_a \, a\\
 c_b\, b&c_c \, c
\end{pmatrix}\;. 
\end{cases}
\ee 
Here, $c_a, c_b,c_c$ and $c_d$ are the Clebsch factors\footnote{
Available Clebsch factors in SU(5) GUTs are, e.g., $c_a, c_b, c_c, c_d \in \{\frac{1}{2},1,\frac{3}{2},3,\frac{9}{2},6,9\}$ and in Pati-Salam models, e.g., $c_a, c_b, c_c, c_d \in \{\frac{3}{4},1,2,3,9\}$. For their viability in supersymmetric scenarios, see e.g.~\cite{Antusch:2009gu}.} which relate the elements in the Yukawa matrices $Y_d$ and $Y_e$. Note that, in SU(5) GUTs, $Y_d$ is related to $Y_e^T$, whereas in Pati-Salam unified theories, $Y_d$ is related directly to $Y_e$.
\end{description}

From the above relations between $Y_d$ and $Y_e$ it is clear that the Clebsch factors play an important role. For a successful model, a consistent set of Clebsch factors, leading also to viable mass relations for the first two families, has to be found. As has been studied recently in \cite{Antusch:2011qg,Marzocca:2011dh}, various combinations of phenomenologically viable Clebsch factors exist which can lead to a comparatively large $\theta^{\text{PMNS}}_{13}$. The prediction $\theta^{\text{PMNS}}_{13} \approx  \theta_C / \sqrt{2}$ only emerges from a subset of these combinations of Clebsches, as we now discuss in the context of SU(5) GUTs and Pati-Salam unified theories.

\subsection{$\theta_{13}^{\text{PMNS}} = \theta_C / \sqrt{2}$ in Pati-Salam Theories}

To obtain the condition on the Clebsch factors, we first note that (in SU(5) GUTs as well as in Pati-Salam models) the 1-2 mixing angle of $Y_d$ is given in leading order in a small mixing approximation by
\be\label{Eq:t12d}
\theta^d_{12} \approx \theta_C \approx \left\vert \frac{b}{c} \right\vert \:.
\ee 
On the other hand, in Pati-Salam unified theories, the 1-2 mixing angle $\theta_{12}^e$ is given by \cite{Antusch:2011qg}
\be
\theta^e_{12} \approx \left\vert\frac{c_b \, b}{c_c \, c}\right\vert   \approx \left\vert\frac{c_b}{c_c}\right\vert   \theta_C \:,
\ee 
where we have made use of the previous equation in the last step. Finally, using Eq.~(\ref{Eq:t13fromt12e}) we obtain
\be\label{Eq:PSt13}
\theta^{\text{PMNS}}_{13} \approx  \frac{\theta_{12}^{\EL} }{\sqrt{2}} 
\approx  \left\vert\frac{c_b}{c_c}\right\vert  \frac{\theta_C}{\sqrt{2}} \:.
\ee
From here we can read off the condition on the Clebsches in Pati-Salam models:

\begin{description}
\item[Condition 4 (PS):] In Pati-Salam unified models, and in general in unified models with a direct GUT relation between $Y_d$ and $Y_e$ (and not as in SU(5) between $Y_d$ and $Y_e^T$) we have to require that the Clebsch factors for the operators generating the 2-2 element and the 1-2 element are equal, i.e.\ 
\be
|c_b| = |c_c| \:.
\ee

\end{description}

In Pati-Salam unified models, this simple additional condition is indeed sufficient and may readily be implemented in flavour models. In such a model, the remaining parameters and Clebsch factors of course have to be chosen to satisfy the phenomenological constraints on the down-type quark and charged lepton sectors. 
We would like to remark at this point that the 1-1 elements of $Y_d$ and $Y_e$ have to be non-zero to yield a viable scenario with the Clebsch factors available in Pati-Salam models \cite{Antusch:2011qg}. Such a texture zero, on the other hand, appears in many models since it leads to the phenomenologically successful GST relation \cite{Gatto:1968ss} and enhances the predictivity of the model. We will see in the next subsection that, in SU(5) GUTs, zero 1-1 elements of $Y_d$ and $Y_e$ are consistent (even favourable) for obtaining the relation $\theta^{\text{PMNS}}_{13} \approx \theta_C / \sqrt{2} $.

\subsection{$\theta_{13}^{\text{PMNS}} = \theta_C / \sqrt{2}$ from SU(5) GUTs}
With $\theta^d_{12}$ given by Eq.~(\ref{Eq:t12d}) and $\theta_{12}^e$ extracted from Eq.~(\ref{Eq:YevsYd}) we have
\be\label{Eq:SU(5)t12d}
\theta^d_{12} \approx \theta_C \approx \left\vert \frac{b}{c}  \right\vert     \quad \mbox{and} \quad  \theta^e_{12} \approx \left\vert\frac{c_a \, a}{c_c \, c}\right\vert  \;,
\ee 
which means that to obtain a link between $\theta_C$ and $\theta_{12}^e$ we want to relate $\left\vert \frac{b}{c}  \right\vert$ to $\left\vert\frac{c_a \, a}{c_c \, c}\right\vert$. This implies that in SU(5) GUTs, in order to establish a simple link between $\theta^{\text{PMNS}}_{13}$ and $\theta_C$, we have to impose at least one further constraint on the structure of the Yukawa matrices.

\begin{description}
\item[Condition 4 (SU(5) GUTs):]  From Eq.~(\ref{Eq:SU(5)t12d}) one can see that one possibility to restore a direct link between $\theta_C$ and $\theta_{12}^e$ is to enforce 
\be
|a| \approx |b| \;
\ee
which immediately translates into
\be\label{Eq:SU5t13}
\theta^e_{12} \approx  \left\vert\frac{c_a \, a}{c_c \, c}\right\vert \approx \left\vert\frac{c_a}{c_c}\right\vert   \theta_C \:
\ee 
and thus to a simple condition 
\be
|c_a| = |c_c| \quad \Rightarrow \quad \theta^{\text{PMNS}}_{13} \approx \frac{\theta_C}{\sqrt{2}} 
\ee 
on the Clebsches in SU(5) GUTs. Two ways to enforce $|a| = |b|$ (or $|a| \approx |b|$ to a good approximation) will be discussed in the following.
\end{description}

\begin{itemize}
\item The first possibility is rather straightforward, namely that the Yukawa matrices $Y_d$ and $Y_e$ are symmetric in the 1-2 submatrix considered above. Although this is not a typical feature in SU(5) GUTs, symmetric matrices may be a consequence of the way that the flavour structure arises out of the breaking of a (non-Abelian) family symmetry and indeed appear in many flavour models. This would imply that
\be
|a| = |b| \quad \mbox{and}\quad |c_a| = |c_b| \; , 
\ee
such that the above condition is satisfied. We note that in order to obtain realistic mass ratios, in the case of symmetric $Y_d$ and $Y_e$, the 1-1 elements have to be non-vanishing and the Clebsch coefficient $c_d$ has to be chosen appropriately. 

\item The second possibility is less straightforward but leads to almost exactly the same prediction. It arises when we impose zero 1-1 elements of $Y_d$ and $Y_e$. The structure of $Y_d$ and $Y_e$ is then quite predictive, since $m_e, m_\mu, m_d, m_s$ and $\theta^d_{12} = \theta_C$ are determined by only three parameters $|a|,|b|$ and $|c|$ and the corresponding Clebsch coefficients. 
In leading order in a small mixing angle approximation, we obtain:
\be \label{Eq:SU(5)LO}
m_s \approx |c| \; , \quad m_\mu \approx |c_c c | \; ,  \quad  
m_d \approx \left\vert\frac{a\, b}{c} \right\vert  \; , \quad 
m_e \approx \left\vert\frac{c_a \, a\: c_b \,b}{c_c \, c}  \right\vert \; ,  \quad  \theta_C \approx  \left\vert\frac{b}{c}\right\vert \; .
\ee
Indeed, this situation has been considered in \cite{Antusch:2011qg} and the relation 
\be
\theta^e_{12} \approx \frac{m_e}{m_\mu} \,\left\vert\frac{c_c}{c_b}\right\vert \, \frac{1}{\theta_C} 
\ee
has been derived. Interestingly, using that \cite{Leutwyler:1996qg}
\be
\frac{m_s}{m_d} \approx 19 \pm 1\; , 
\ee
means that to quite good precision we can set\footnote{This experimentally justified relation is also known as the GST relation \cite{Gatto:1968ss}.} 
\be 
\theta_C \approx \sqrt{\frac{1}{19}} \approx \sqrt{\frac{m_d}{m_s}}\;.
\ee 
We can now easily see that also this constraint (i.e.\ zero 1-1 elements of $Y_d$ and $Y_e$) enforces $|a| \approx |b|$ to a very good approximation: From Eq.~(\ref{Eq:SU(5)LO}) we obtain for instance
\be
\frac{m_d}{m_s} \approx \left\vert\frac{a\, b}{c^2} \right\vert \approx  \left\vert\frac{a}{b} \right\vert   \, \theta_C^2 \;,
\ee
which leads to the approximate relation $|a| = |b|$. Although less direct than in the first case, the same connection to $\theta_C$, i.e.\ $\theta^e_{12} \approx  \left\vert\frac{c_a}{c_c}\right\vert   \theta_C$, is found and thus $\theta^{\text{PMNS}}_{13} \approx \theta_C / \sqrt{2}$ is realised.

In this context it should be noted that in addition to $|c_a| = |c_c|$, also $c_b$ has to be chosen properly for consistent quark-lepton mass ratios for the first two families \cite{Antusch:2011qg,Marzocca:2011dh}. In SU(5) GUTs with zero 1-1 elements in $Y_d$ and $Y_e$, only one combination of Clebsches discussed in \cite{Antusch:2011qg,Marzocca:2011dh} results in $\theta^{\text{PMNS}}_{13} \approx \theta_C / \sqrt{2}$, while being consistent with phenomenological constraints (including $m_s/m_d \approx 19$), namely  
\be
c_c = c_a = 6  \; , \quad c_b = \frac{1}{2} \; .
\ee
\end{itemize}

\section{Corrections to $\theta_{13}^{\text{PMNS}} = \theta_C / \sqrt{2}$}
\label{sec:corr}
As we have discussed above, the relation $\theta_{13}^{\text{PMNS}} \approx \theta_C / \sqrt{2}$ can emerge under four simple conditions from GUTs. However, it is important to emphasize again that the relation is not expected to hold exactly, as we now discuss. 


\begin{itemize}

\item {\bf Corrections due to small mixing approximation} 

In section \ref{sec:2} we omitted higher order terms in a small mixing angle expansion to keep our discussion as simple as possible. The error introduced by this simplification depends on the structure of the mass matrices and on the Clebsch factors and can easily be computed. Let us consider, for instance, the example of SU(5) GUTs with zero 1-1 elements in $Y_d$ and $Y_e$ and Clebsch factors $ c_c = c_a = 6$,  $c_b = \frac{1}{2}$, i.e. 
\be
Y_d = \begin{pmatrix} 
0&b\\
a&c
\end{pmatrix}\,,\;\;
Y_e = 
\begin{pmatrix} 
0& c_a \, a\\
 c_b\, b&c_c \, c
\end{pmatrix}\;. 
\ee
Beyond leading order approximation, we obtain $\theta_{12}^e=13.8^\circ$ by fitting the experimental values of $m_e/m_\mu$ and $\theta_C$ to the results of an exact diagonalisation. Instead of $\theta_{13}^{\text{PMNS}} = \theta_C / \sqrt{2} = 9.2^\circ$, the more precise GUT scale prediction is given by 
\be
\mbox{SU(5) with $d=0$, $c_c = c_a = 6,  \:c_b = \frac{1}{2}$ and $\theta_{12}^\text{d} = \theta_C$:} \qquad \theta_{13}^{\text{PMNS}} = 9.8^\circ\;.
\ee

\item {\bf Corrections from $\theta_{12}^d\neq\theta_C$}

In explicit GUT models of flavour, the condition $\theta_{12}^d=\theta_C$ discussed in section \ref{sec:1} may not be exactly fulfilled. Without a specific model, this is a source of theoretical uncertainty. Within a specific model, such a deviation will result in a modified prediction for $\theta_{13}^{\text{PMNS}}$ (which still belongs to  $\theta_{13}^{\text{PMNS}} \approx \theta_C / \sqrt{2}$). 
This may open up the possibility to distinguish explicit models with a future more precise $\theta_{13}^{\text{PMNS}}$ measurement.  
For example, let us consider deviations from $ \theta_{12}^\text{d} = \theta_C$ which arise in classes of models where $\theta_{13}^u$, $\theta_{13}^d\approx 0$, which leads to $\theta_{12}^d\approx 12.0^\circ\pm0.3^\circ$ via the quark mixing sum rule $\theta_{12}^d\approx\Big |\theta_{12}^\text{CKM}-\frac{\theta_{13}^\text{CKM}}{\theta_{23}^\text{CKM}}e^{-i\delta^\text{CKM}}\Big |$ (cf.~\cite{Antusch:2009hq}). Let us consider a Pati-Salam model with symmetric Yukawa matrices and Clebsch factors $c_d=9$ and $c_b=c_c=- 3$, i.e.
\be
Y_d = \begin{pmatrix} 
d&b\\
b&c
\end{pmatrix}\,,\;\;
Y_e = 
\begin{pmatrix} 
c_d\, d& c_b \, b\\
 c_b\, b&c_c \, c
\end{pmatrix}\;. 
\ee
Fitting the experimental values to the results of an exact diagonalization
we obtain the modified GUT scale prediction (assuming $\theta_{23}^\text{PMNS}=45^\circ$) 
\be
\mbox{PS with $\theta_{13}^u \approx \theta_{13}^d\approx 0$, $a=b$, $c_d=9$ and $c_b=c_c= - 3$:} \qquad \theta_{13}^{\text{PMNS}} = 8.6^\circ\;.
\ee

\newpage

\item {\bf Corrections due to deviations from $\theta_{23}^\text{PMNS}=45^\circ$}

Another simplification we have used is $\theta_{23}^\text{PMNS} = 45^\circ$, and we have commented already in section 2 that the more accurate relation which, for $\theta_{23}^\text{PMNS} \not= 45^\circ$, should be applied instead of $\theta^{\text{PMNS}}_{13} \approx \frac{1}{\sqrt{2}} \theta_{12}^{\EL}$,  is 
\be\label{eq:rel_with_s23}
\theta^{\text{PMNS}}_{13} \approx \theta_{12}^{\EL}s_{23}^{\text{PMNS}} \;,
\ee 
up to correction of ${\cal O}((\theta^{\text{PMNS}}_{13})^3)$. With the measured range for $\theta_{23}^\text{PMNS}$, this already introduces an uncertainty of about 6\%. In this respect, it is interesting to note that some recent global fits see hints for $\theta_{23}^\text{PMNS} < 45^\circ$. For instance the best fit of \cite{Fogli:2011qn} is given by $\theta_{23}^\text{PMNS}=40.4^\circ$ with a one sigma range of $[38.6^\circ,45.0^\circ]$. If such hints get confirmed, this would imply a lower predicted value of $\theta_{13}^{\text{PMNS}}$. For example, for $\theta_{23}^\text{PMNS}=40.4^\circ$, the deviation from the relation assuming maximal mixing is $\theta_C s_{23}^\text{PMNS} - \theta_C / \sqrt{2}=-0.8^\circ$. 
An improved experimental accuracy for $\theta_{23}^\text{PMNS}$ would be important for more precise $\theta_{13}^{\text{PMNS}}$ predictions.

\item {\bf Corrections due to RG running}

Let us now discuss the impact of RG corrections. For simplicity, we will focus on the case of a strongly hierarchical neutrino mass spectrum with normal (NH) 
or inverse (IH) mass ordering (with $m_1=0$ and $m_3=0$, respectively) in the MSSM. We will base our discussion on the relation 
\be\label{eq:rel_with_s23}
\theta^{\text{PMNS}}_{13} = \theta_C \, s_{23}^{\text{PMNS}}   \;,
\ee
which allows for general $\theta^{\text{PMNS}}_{23}$. Since the relation is defined at the GUT scale $M_\text{GUT}$, the strategy is to run the measured values of $\theta_C$ and $\theta^{\text{PMNS}}_{23}$ up to $M_\text{GUT}$ to determine $\theta^{\text{PMNS}}_{13} |_{M_\text{GUT}}$. Then, the RG running of $\theta^{\text{PMNS}}_{13} |_{M_\text{GUT}}$ down to the electroweak scale $M_\text{EW}$ is performed to obtain $\theta^{\text{PMNS}}_{13} |_{M_\text{EW}}$, the parameter measurable in experiments.  

Since the running of $\theta_C$ is known to be tiny, we will neglect it in the following. Using the analytical results of \cite{Antusch:2003kp}, one can estimate $\Delta s_{23}^\text{PMNS}\equiv~s_{23}^\text{PMNS}\vert_{M_\text{EW}}-
s_{23}^\text{PMNS}\vert_{M_\text{GUT}}$ in leading logarithmic approximation and in leading (zeroth) order in $\theta^{\text{PMNS}}_{13} $:
 \begin{align}
\label{eq:36}\mbox{NH:}\qquad
\Delta s_{23}^\text{PMNS}\approx & \frac{(y_\tau^\text{SM})^2(1+\tan^2\beta)}{16\pi^2}\ln\left(\frac{M_\text{GUT}}{M_\text{EW}}\right)(c_{23}^\text{PMNS})^2s_{23}^\text{PMNS}\;,\\
\mbox{IH:}\qquad
\Delta s_{23}^\text{PMNS}\approx &-\frac{(y_\tau^\text{SM})^2(1+\tan^2\beta) }{16 \pi^2}\ln\left(\frac{M_\text{GUT}}{M_\text{EW}}\right)(c_{23}^\text{PMNS})^2s_{23}^\text{PMNS}\;.
\end{align}
For the running of $\theta^{\text{PMNS}}_{13}$, we also include the terms of ${\cal O}(\theta^{\text{PMNS}}_{13})$, and obtain for $\Delta\theta_{13}^\text{PMNS}\equiv \theta_{13}^\text{PMNS}\vert_{M_\text{EW}}-\theta_{13}^\text{PMNS}\vert_{M_\text{GUT}}$ in leading order in $\theta^{\text{PMNS}}_{13} $ and $\sqrt{|\Delta m^2_{21}/\Delta m^2_{31}|}$ (with $\Delta m^2_{ij} = m_i^2 - m_j^2$ being the neutrino mass squared differences) \cite{Antusch:2003kp}:
 \begin{align}
\mbox{NH:}\qquad
\Delta\theta_{13}^\text{PMNS}\approx & \frac{(y_\tau^\text{SM})^2(1+\tan^2\beta)}{16\pi^2}\ln\left(\frac{M_\text{GUT}}{M_\text{EW}}\right)\left((c_{23}^\text{PMNS})^2\theta_{13}^\text{PMNS}\vphantom{\frac{m2}{m3}}\right. \nonumber\\
& +\left. 2\frac{m_2}{m_3}\cos(\delta^\text{PMNS}-\varphi_2^\text{PMNS})c_{12}^\text{PMNS}s_{12}^\text{PMNS}
c_{23}^\text{PMNS}s_{23}^\text{PMNS}\right) \;,\\
\label{eq:39} \mbox{IH:}\qquad
\Delta\theta_{13}^\text{PMNS} \approx &-\frac{(y_\tau^\text{SM})^2(1+\tan^2\beta) }{16 \pi^2}\ln\left(\frac{M_\text{GUT}}{M_\text{EW}}\right)(c_{23}^\text{PMNS})^2\theta_{13}^\text{PMNS}\;.
\end{align}

We can now calculate $\theta^{\text{PMNS}}_{13} |_{{M_\text{EW}}}$, based on the measured values of $\theta_C$ and $\theta^{\text{PMNS}}_{23}\vert_{M_\text{EW}}$, but using the GUT scale relation of Eq.~(\ref{eq:rel_with_s23}):
\bea\label{eq:40}
\theta^{\text{PMNS}}_{13} |_{M_\text{EW}} =  \theta_C \,  s_{23}^\text{PMNS}\vert_{M_\text{GUT}} +  \Delta\theta_{13}^\text{PMNS} = \theta_C \,  (s_{23}^\text{PMNS}\vert_{M_\text{EW}} - \Delta s_{23}^\text{PMNS}) +  \Delta\theta_{13}^\text{PMNS}
\eea 
One can see that for the IH case the terms $\Delta\theta_{13}^\text{PMNS}$ and $ \theta_C \, \Delta s_{23}^\text{PMNS}$ cancel each other at leading order, while for NH only the term proportional to the neutrino mass ratio $\frac{m_2}{m_3}$ remains. Plugging in the experimental values of the mixing parameters and $y_\tau^\text{SM} \approx 0.01$ we obtain the following estimate for the effects of RG running:
\begin{align}
\mbox{NH:}\qquad\theta^{\text{PMNS}}_{13} |_{M_\text{EW}} &\approx \theta_C \,  s_{23}^\text{PMNS}\vert_{M_\text{EW}} + 0.2^\circ \cos(\delta^\text{PMNS}-\varphi_2^\text{PMNS})\left(\frac{\tan\beta}{50}\right)^2 \;,\\
\mbox{IH:}\qquad \theta^{\text{PMNS}}_{13} |_{M_\text{EW}} &\approx  \theta_C \,  s_{23}^\text{PMNS}\vert_{M_\text{EW}} \;,
\end{align}  
where the corrections in the IH case from the next order terms are estimated about ${\cal O}(0.05^\circ) \left(\frac{\tan\beta}{50}\right)^2$. It is thus a rather good approximation to evaluate Eq.~(\ref{eq:rel_with_s23}) with the measured parameters at low energies. For the IH case, the prediction is remarkably insensitive to RG effects.

We remark that the above treatment does not include possible effects from neutrino Yukawa couplings, which may contribute above the mass thresholds of the right-handed neutrinos in type I seesaw models. These effects are more model-dependent. They can be estimated using the analytical formulae of \cite{Antusch:2005gp} or calculated numerically, e.g.\ using the REAP package (introduced in \cite{Antusch:2005gp}). For hierarchical neutrino Yukawa matrices (dominated by the 3-3 element $y_{33}$) one can obtain an estimate for the additional correction by simply replacing $(y_\tau^\text{SM})^2(1+\tan^2\beta) \ln (M_\text{GUT}/M_\text{EW})$ in the above equations by $y_{33}^2 \ln (M_\text{GUT}/M_\text{R3})$, where $M_\text{R3}$ is the mass of the corresponding right-handed neutrino.

\newpage

\item {\bf Corrections from canonical normalisation}

Furthermore, also effects from canonical normalisation (CN) of the kinetic terms can lead to deviations from $\theta_{13}^{\text{PMNS}} = \theta_C \, s_{23}^{\text{PMNS}} $. The resulting corrections can be estimated with analogous formulae as used above for the RG corrections, when they dominantly result from third family effects, as has been discussed in \cite{CN1,CN2}. In this case, the CN effects can be parameterized by a single (model-dependent) parameter $\eta^\text{CN}$ (for the definition of $\eta^\text{CN}$, see \cite{CN1,CN2}) and the formulae (\ref{eq:36}) - (\ref{eq:39}) can be used, replacing $(y_\tau^\text{SM})^2(1+\tan^2\beta) \ln (M_\text{GUT}/M_\text{EW})$ by $8\pi^2 \eta^\text{CN}$ and interpreting the RG change now as a correction from canonical normalisation. As for RG corrections, the relation is also quite insensitive to CN corrections in the IH case. The size of $\eta^\text{CN}$, on the other hand, is highly model-dependent. The CN effects on $\theta_{13}^{\text{PMNS}} = \theta_C \, s_{23}^{\text{PMNS}} $ in the NH case may therefore well be of the order of the RG corrections (or even larger) in some specific models. On the other hand, there are classes of models where the CN corrections are negligible (cf.\ \cite{CN1}). 

\end{itemize}

In summary, we conclude that the uncertainties for the relation $\theta_{13}^{\text{PMNS}} = \theta_C / \sqrt{2} \approx 9.2^\circ$ amount to less than about ${\cal O}(10\%)$, which is of the order of the current experimental uncertainty. On the other hand, in an explicit GUT model of flavour where the corrections can be calculated, a more precise prediction can be obtained. It is important to note that Daya Bay has already presented new preliminary results at Neutrino 2012 and ICHEP 2012 \cite{talks}, where $\theta_{13}^\text{PMNS}=8.7^\circ\pm0.8^\circ$, and the experiment further aims at an uncertainty of only $\pm 0.25^\circ$. This high precision would be of the order of (or better than) the theory corrections mentioned above and may help to discriminate between specific flavour models. We emphasize that towards this end a careful model analysis, including all these corrections, will be required.

\section{The underlying neutrino mixing pattern in the light of $\theta_{13}^{\text{PMNS}} = \theta_C / \sqrt{2}$}
\label{sec:sumrule}

Under the assumption $\theta_{13}^\nu$, $\theta_{13}^e\ll\theta_C$ (cf.\ condition 1), the mixing angle $\theta_{12}^\nu$ is related to $\theta_{13}^{\text{PMNS}}$ and $\theta_{12}^{\text{PMNS}}$ by the lepton mixing sum rule\footnote{The sum rule becomes $\theta_{12}^{\text{PMNS}}  - \theta_{13}^{\text{PMNS}} \cot (\theta_{23}^\text{PMNS}) \cos (\delta^{\text{PMNS}} ) \approx \theta_{12}^\nu$ if $\theta_{23}^\text{PMNS}$ deviates from maximal mixing \cite{Antusch:2005kw,sumrule}. RG corrections to the sum rule have been discussed in \cite{Boudjemaa:2008jf}.\label{footnote_sum_rule}} \cite{Antusch:2005kw,sumrule}
\be\label{Eq:SumRule}
\theta_{12}^{\text{PMNS}}  - \theta_{13}^{\text{PMNS}}  \cos (\delta^{\text{PMNS}} ) \approx \theta_{12}^\nu\; .
\ee
The relation $\theta_{13}^{\text{PMNS}} \approx \theta_C / \sqrt{2}$ has interesting consequences: 
\begin{itemize}
\item Assuming tri-bimaximal mixing in the neutrino sector \cite{TBM}, i.e.\ $\sin(\theta_{12}^\nu)=1/\sqrt{3}$, the sum rule becomes
\be
\theta_{12}^{\text{PMNS}}  -  \frac{\theta_C}{\sqrt{2}} \cos (\delta^{\text{PMNS}} ) \approx \arcsin\left(\frac{1}{\sqrt{3}}\right)\;.
\ee
This implies that  $\theta_{13}^{\text{PMNS}} \approx \theta_C / \sqrt{2}$ can only be consistent with tri-bimaximal neutrino mixing for $\delta^{\text{PMNS}}\approx \pm 90^\circ$.\footnote{Such specific predictions for $\delta^{\text{PMNS}}$ may emerge from flavour models with $\mathbb{Z}_2$ or $\mathbb{Z}_4$ shaping symmetries to explain a right-angled \text{\text{CKM}} unitarity triangle (with $\alpha \approx 90^\circ$), as discussed recently in \cite{Antusch:2011sx}.}
\item Another possibility would be bi-maximal neutrino mixing \cite{Barger:1998ta}, i.e.\ $\theta_{12}^\nu = 45^\circ$. Then the sum rule reads
\be
\label{Eq:32}
\theta_{12}^{\text{PMNS}}  -  \frac{\theta_C}{\sqrt{2}} \cos (\delta^{\text{PMNS}} ) \approx 45^\circ \;,
\ee
which differs by the factors $-\cos(\delta^\text{PMNS})$ and $1/\sqrt{2}$ from the common form of the QLC relation.
For $\delta^\text{PMNS}\approx 180^\circ$, Eq.~(\ref{Eq:32}) agrees reasonably well with the present experimental data.
\end{itemize}

To distinguish the above mentioned two possibilities, it may be interesting to note that, while models with tri-bimaximal neutrino mixing seem to be associated with a normal neutrino mass hierarchy, models with bi-maximal mixing rather tend to emerge in the context of models with inverse neutrino mass hierarchy, for instance with an appropriate U(1) symmetry (like L$_e$ - L$_\mu$ - L$_\tau$ \cite{Barbieri:1998mq}).   

\begin{figure}[h!]
\center
\includegraphics[scale=.7]{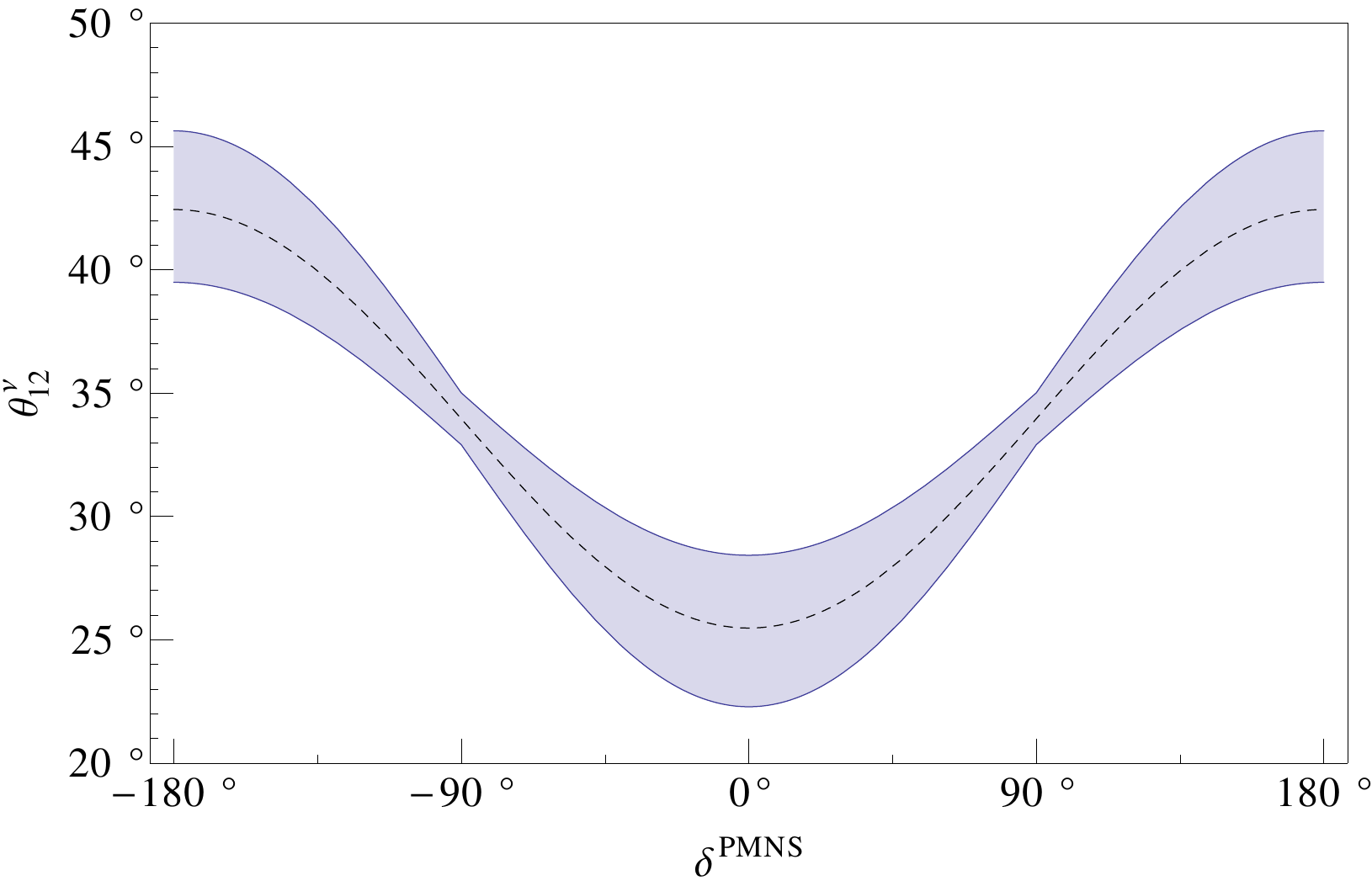}
\caption{Using the lepton mixing sum rule of Eq.\ (\ref{Eq:SumRule}), a measurement of $\delta^\text{PMNS}$ allows to determine $\theta_{12}^\nu$ (under the condition $\theta_{13}^\nu$, $\theta_{13}^e\ll\theta_C$). The shaded region is obtained from inserting the $1\sigma$ uncertainties of $\theta_{12}^\text{PMNS}$, $\theta_{13}^\text{PMNS}$ and $\theta_{23}^\text{PMNS}$ into the sum rule, i.e.\ without assuming maximal $\theta_{23}^\text{PMNS}$.}
\label{figuresumrule}
\end{figure}

Finally, it should be emphasized that neither tri-bimaximal nor bi-maximal neutrino mixing are required for obtaining the prediction $\theta_{13}^{\text{PMNS}} \approx \theta_C / \sqrt{2}$ from GUTs. As stated above, in the neutrino sector it is only required that $\theta^\nu_{13} \ll \theta_C$. In this sense, a measurement of $\delta^{\text{PMNS}}$ may also be viewed as ``reconstructing'' the value of $\theta_{12}^\nu$, as shown in Fig.\ \ref{figuresumrule}.

In general, using the neutrino mixing sum rule of Eq.~(\ref{Eq:SumRule}), a precise measurement of $\theta_{12}^{\text{PMNS}} ,\theta_{13}^{\text{PMNS}} $ and $\delta^{\text{PMNS}} $ may hint at a specific neutrino mixing pattern. The measurement of $\theta_{13}^{\text{PMNS}} \approx \theta_C / \sqrt{2}$ provides a valuable input in this context.

\section{Summary and conclusions}
\label{sec:sum}
Motivated by the recent measurements of the leptonic mixing angle 
\be
\theta^{\text{PMNS}}_{13} = 8.8^\circ \pm 1.0^\circ \;,
\ee
which are consistent with $\theta^{\text{PMNS}}_{13} = \theta_C / \sqrt{2}$, we have discussed how this relation can emerge in Grand Unified Theories.

The key towards realising this relation is that in GUTs the down-type quark Yukawa matrix $Y_d$ and the charged lepton Yukawa matrix $Y_e$ are generated from the same set of GUT operators, which implies that the resulting matrix elements are equal up to group theoretical Clebsch factors. This can lead to the link 
\be 
\theta^e_{12} \approx \theta_C
\ee 
between quark and lepton mixing, which can then induce  
\be 
\theta^{\text{PMNS}}_{13} \approx \frac{\theta_C}{\sqrt{2}}
\ee 
from the charged lepton contribution $\theta^e_{12}$ to the leptonic mixing matrix. This connection is illustrated in Fig.\ \ref{picture}, which also shows the simple conditions we find for obtaining the relation $\theta^{\text{PMNS}}_{13} \approx \theta_C / \sqrt{2}$ in SU(5) GUTs and Pati-Salam models:

\begin{figure}
 \center
 \begin{tikzpicture}[
      scale=0.5,
      circ/.style={thick,circle},
      rec/.style={thick,rectangle},
      trans/.style={thick,<->,shorten >=2pt,shorten <=2pt,>=stealth},
    ]
    \draw[circ] (0,8) node[style={fill=black!20},scale=1.2] {$Y_u$};
    \draw[circ] (8,8) node[style={fill=black!20},scale=1.2] {$Y_d$};
    \draw[circ] (8,0) node[style={fill=black!20},scale=1.2] {$Y_e$};
    \draw[circ] (0,0) node[style={fill=black!20},scale=1.2] {$m_\nu$};
    
    \draw[rec] (-3,0) node[rectangle,fill=white,draw,scale=1,left] {Condition 1};
    \draw[rec] (-3,8) node[rectangle,fill=white,draw,scale=1,left] {Condition 2};
    \draw[rec] (11,4) node[rectangle,fill=white,draw,scale=1,right] {Conditions 3 \& 4};

    \draw[trans] (2,0) -- (6,0) node[midway,above,scale=1] {$U_\text{PMNS}$};
    \draw[trans] (2,8) -- (6,8) node[midway,above,scale=1] {$U_\text{CKM}$};
    \draw[trans] (8,6) -- (8,2) node[midway,below,scale=1,rotate=90] {GUTs};
    \end{tikzpicture}
\caption{Under conditions 1 to 4, the relation $\theta_{13}^{\text{PMNS}} \approx \theta_C / \sqrt{2}$ is obtained by linking the charged lepton and quark sectors via GUTs.}
\label{picture}
\end{figure}
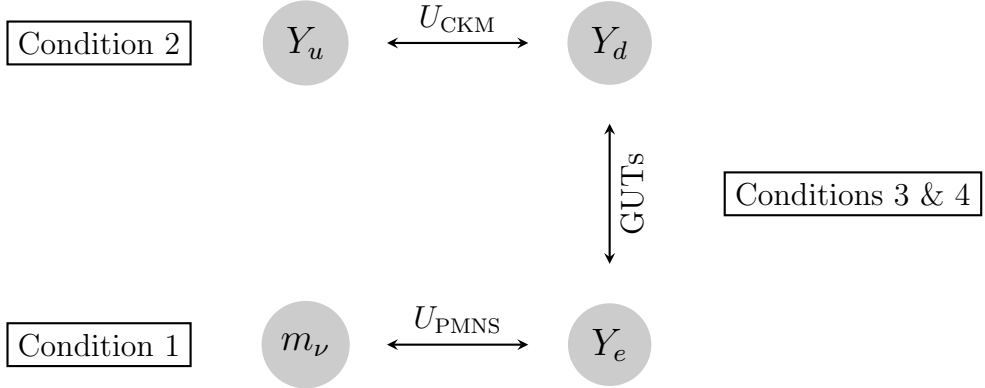

\begin{description}  
\item[Condition 1:] The 1-3 mixings in $m_\nu$ and $Y_e$ should be small, i.e.\ much smaller than $\theta_C$,
\be
\theta_{13}^\nu \approx 0 \; , \quad \theta_{13}^e \approx 0\;.
\ee
\item[Condition 2:] The 1-2 mixing in $Y_d$ should be equal to the Cabibbo angle to a good approximation,
\be
\theta^d_{12} \approx \theta_C \: ,
\ee
which is, for example, automatically satisfied if $\theta_{12}^u\ll\theta_{12}^d$.
\item[Condition 3:] The relevant elements of $Y_d$ and $Y_e$ have to be generated dominantly by one single GUT operator (or by operators with the same group theoretical Clebsch factors). 
\item[Condition 4:] Two of the Clebsch factors should be equal, i.e.\ $|c_b| = |c_c|$ in Pati-Salam models and $|c_a| = |c_c|$ in SU(5) GUTs. In SU(5) GUTs, one additional constraint has to be imposed on the structure of the mass matrices, such as a symmetry in the 1-2 submatrix or zero 1-1 elements of $Y_d$ and $Y_e$. 
\end{description}

We have also analysed various corrections to $\theta_{13}^\text{PMNS}=\theta_C / \sqrt{2}$, e.g.\ from renormalization group running, from next-to-leading terms in the small angle expansion and from a deviation of $\theta_{23}^\text{PMNS}$ from $45^\circ$. Without specifying an explicit model we estimate the corrections to be less than about $\mathcal{O}(10\%)$, i.e.\ of the order of the present experimental uncertainty. 
In an explicit GUT model of flavour, on the other hand, the theoretical corrections may be calculated and a more precise prediction can be obtained, so that future more precise measurements of $\theta_{13}^\text{PMNS}$ and $\theta_{23}^\text{PMNS}$ may allow to discriminate such specific flavour models with $\theta_{13}^\text{PMNS} \approx \theta_C / \sqrt{2}$. 

In a model-independent way, we have furthermore discussed how, using lepton mixing sum rules, specific mixing patterns for the neutrino sector can be distinguished by their different predictions for the Dirac CP phase $\delta^{\text{PMNS}}$ (cf.\ Fig.~\ref{figuresumrule}).  A measurement of $\delta^{\text{PMNS}}$ may also be viewed as ``reconstructing'' the value of $\theta_{12}^\nu$, provided that $\theta_{13}^\nu$, $\theta_{13}^e\ll\theta_C$ (cf. condition 1) holds. In this context we have also emphasized that no specific neutrino mixing pattern has to be assumed for obtaining $\theta_{13}^{\text{PMNS}} \approx \theta_C / \sqrt{2}$ from GUTs. In the neutrino sector it is sufficient to require that $\theta_{13}^\nu \ll \theta_C$. 

In summary, we have shown that the relation $\theta_{13}^{\text{PMNS}} \approx \theta_C / \sqrt{2}$ may readily be obtained from SU(5) GUTs and unified theories with Pati-Salam structure, if some simple conditions are satisfied.

\section*{Acknowledgements}
This project was supported by the Swiss National Science Foundation. S.~A.\ acknowledges partial support by the DFG cluster of excellence ``Origin and Structure of the Universe''.

\newpage

\section*{Note added}
While this paper was finalized, \cite{TBC} appeared which also discusses strategies for obtaining $\theta_{13}^{\text{PMNS}} \approx \theta_C / \sqrt{2}$. We go beyond this paper, for instance by including the case of SU(5) GUTs, which differs substantially from the case of Pati-Salam models.

\end{document}